\documentclass[a4paper,10pt]{aa}
\usepackage{graphics}
\usepackage{natbib}
\begin{document}
\newcommand{\Msun}{M$_{\sun}$\ }
\newcommand{\Lsun}{L$_{\sun}$\ }
\newcommand{\Mstar}{M$_{*}$\ }
\newcommand{\Lstar}{L$_{*}$\ }
\newcommand{\Ha}{H{$\alpha$}\ }
\newcommand{\LH}{LkH{$\alpha$}\,233\ }
\newcommand{\Lk}{LkH{$\alpha$}\,234\ }
\newcommand{\AC}{Associated Cloud}
\newcommand{\FL}{Flow length }
\newcommand{\CS}{Cloud size$^a$ }
\newcommand{\tf}{$\theta$$_{flow}$$^b$ }
\newcommand{\1}{LkH{$\alpha$} 198\ }
\newcommand{\2}{1548C27\,IRS\,1}
\newcommand{\3}{IRAS\,19395+2313 }
\newcommand{\4}{IRAS\,18162\,-\,2048$^c$ }
\newcommand{\5}{HH\,354\,IRS }
\newcommand{\6}{HH\,1/2 VLA\,1$^e$ }
\newcommand{\7}{HH\,34\,IRS$^e$ }
\newcommand{\8}{IRAS\,05491+0247$^{d,e}$ }
\newcommand{\9}{PV Cephei}
\newcommand{\yr}{$\times$10$^4$ yr }
\newcommand{\aar}{A{\&}A Rev.\ }
\newcommand{\aas}{A{\&}AS\ }
\newcommand{\annrev}{ARA{\&}A\ }

\title{Parsec\,-\,scale Herbig-Haro Outflows from Intermediate Mass Stars}
\author{F. McGroarty, \inst{1}
T.P. Ray \inst{1}
\and J. Bally \inst{2}}
\offprints{F. McGroarty, fmcg@cp.dias.ie}
\institute{Dublin Institute for Advanced Studies, 5 Merrion Square, 
Dublin 2, Ireland
\and Department of Astrophysical and Planetary Sciences and Center for 
Astrophysics and Space Astronomy, University of Colorado, Campus Box 389, 
Boulder, CO 80309-0389, USA}
\date{Received date ;accepted date}

\abstract{
While there are many parsec\,-\,scale Herbig-Haro (HH) outflows known to be 
driven by low\,-\,mass young stars, few are associated with their 
intermediate mass counterparts. Here we present the discovery of five such 
bipolar outflows. Of these, LkH$\alpha$\,198, 1548C27\,IRS\,1, \LH and 
\Lk were previously known to possess small-scale HH flows, while no such 
activity was observed before near IRAS\,19395+2313. The largest 
of the newly discovered outflows are seen in the vicinity of \Lk and 
1548C27\,IRS\,1, 
and stretch (in projection) 8\,pc and 7.5\,pc respectively. \LH which
was previously known to power a spectroscopically detected small-scale 
($\leq$10\arcsec) jet is now seen to drive a 3\,pc outflow and 
LkH$\alpha$\,198 is shown here to power a 2\,pc outflow. Two HH objects in 
the vicinity of IRAS\,19395+2313 lead us to suggest that it may also be 
responsible for a 5\,pc outflow. In total, 27 new HH 
objects/complexes were discovered. Examination of 
these parsec\,-\,scale outflows show that they have similar lengths, 
morphologies, and dynamical timescales as those from 
low\,-\,mass sources. Many appear to have blown out of the parent cloud, 
suggesting that their total lengths are much greater than optically observed.
The degree of collimation of these outflows is similar to those from low\,-\,mass 
sources suggesting that the transition to more poorly\,-\,collimated outflows must 
occur at higher masses than the sources observed here.

\keywords{ISM: Herbig-Haro objects --- jets and outflows, 
Stars: pre-main sequence --- formation, Individual --- LkH$\alpha$\,198, 
1548C27 IRS1, \LH, \Lk}
}
\maketitle

\section{Introduction}

\indent HH objects are the shock\,-\,excited nebulous tracers of outflows from 
pre\,-\,main sequence stars. In many cases these outflows are collimated in 
the form of highly supersonic jets the existence of which appear to be 
intrinsicly linked to accretion by the underlying young stellar object 
(YSO) \citep{Hartigan95}. Most known optical jets have low mass ($\sim$ 1\Msun) 
sources: either the  embedded (IRAS Class I) counterparts of classical T\,Tauri 
stars or classical T\,Tauri stars themselves \citep{Reipurth97}.

Turning to higher mass YSOs ($\verb+>+$ 10\Msun), such as those driving the 
Orion OMC\,1 or Cepheus A outflows, one sees a very different picture 
\citep{Allen93, ODell97, Hartigan20}. Their outflows, although highly energetic, 
often appear poorly collimated and more chaotic \citep{Reipurth01} than their 
low mass counterparts. This transformation begs two obvious inter-related 
questions: at what point does the transition occur and is it a 
smooth function of mass? To answer these questions one must examine outflows 
from YSOs of mass greater than $\sim$2\,\Msun$\!\!$.

Optical outflows have been observed from a number of intermediate\,-\,mass 
(2\,\Msun$\leq$\,\Mstar$\leq$\,10\,\Msun$\!\!$) YSOs, for example R\,Mon, 
\Lk$\!\!$, and AFGL\,4029 (Ray et al., 1990 and references therein). Such 
stars, where optically visible, are known as Herbig Ae/Be stars (HAEBES) 
although their embedded counterparts have also been seen. However optical 
outflows from these YSOs are rare and there are a number of reasons for 
this. The initial mass function favours the production of low mass stars 
and therefore intermediate mass YSOs tend to be found at relatively large 
distances. More massive stars also have a faster evolutionary timescale 
-- i.e.\ they evolve more quickly than low\,-\,mass YSOs and so their outflow 
phase is shorter. This would also make their outflows more difficult to detect. 
Another contributing factor could be that massive stars tend to be more obscured : 
intermediate and massive stars tend to be surrounded by large amounts of 
circumstellar gas and dust, making it harder to find an outflow at visual 
wavelengths. Finally there may well have been a historical bias towards 
studying outflows from low\,-\,mass stars. The situation however has changed 
in recent years as more and more studies focus on higher mass YSOs.

With these ideas in mind, we have investigated the occurrence of large-scale 
outflows from intermediate mass stars. By large scale we mean those stretching 
several parsecs e.g.\ the PV Cephei outflow at 2.6 pc 
\citep{Gomez97,Reipurth97}, the HH\,80/81 5.3 pc outflow \citep{Marti93} and 
the HH\,354 outflow at 2.4 pc \citep{Reipurth97}. We emphasise that these 
outflows have vastly longer associated timescales than those of 
``traditionally'' observed flows.  A small\,-\,scale HH jet 
close to its source has a dynamical timescale of only a few hundred years, 
whereas the HH objects in these parsec\,-\,scale flows trace mass ejection 
over tens of thousands of years. They are, in effect, fossil records of 
the mass\,-\,loss histories of their parent star.

Newly detected parsec\,-\,scale outflows around five intermediate\,-\,mass 
young stars are discussed here (see also Table\ \ref{HHPositions} and 
Table\ \ref{parameters}). Of these, LkH$\alpha$\,198, \LH and \Lk are 
of spectral type A and 1548C27 is A7\,-\,F0. All of these stars were known to 
possess small-scale optical outflows. The one optically invisible YSO in our 
sample, IRAS\,19395+2313, was not previously known to drive any outflow.  

Section 2 describes how we made our observations and in Section 3 
we present our newly discovered large-scale flows. The implications of  
our findings are discussed in Section 4 and our conclusions are drawn 
in Section 5.

\section{Observations}

\indent To carry out our survey we used the Wide Field Camera (WFC) at the 
prime focus of the 2.5m Isaac Newton Telescope at El Observatorio del Roque de 
los Muchachos (La Palma, Canary Islands). The WFC consists of four 
thin\,-\,coated EEV CCDs each with 2048\,x\,4100 15$\mu$m$^2$ pixels. One pixel 
projects to 0\farcs33 on the sky. Three of the CCDs are positioned from north
to south with their long axes adjoining. The fourth is attached to the west to 
form a square mosaic (34\farcm2 wide) with its 
northwestern corner missing.

Our images were taken on nights between the 13$^{\rm th}$ and the 
21$^{\rm st}$ of July 1998. Seeing was moderate at 1\farcs15\,--\,1\farcs5 
as measured from the images. HH objects were 
identified using a number of narrowband emission line filters:  
H$\alpha$($\lambda_c$ = 6568\AA, $\Delta\lambda$(FWHM) = 95\AA ), 
[SII]($\lambda_c$ = 6725\AA, $\Delta\lambda$(FWHM) = 80\AA ) and 
[OIII]($\lambda_c$ = 5008\AA, $\Delta\lambda$ (FWHM) = 100\AA). 
To distinguish HH emission from reflection nebulosity, we also took broadband 
images in V and I. Exposure times for the narrowband and broadband images were 
typically 30 and 10 minutes respectively. 
The data was reduced using standard IRAF reduction procedures. 

\begin{table} 
\begin{tabular}{llll}
\hline \hline
Object       &Source            &$\alpha$(J2000)       &$\delta$(J2000)         \\ \hline
HH\,800      &UNKNOWN           &00$^h$11$^m$02.0$^s$  &+58\degr 55$'$04$''$    \\ 
HH\,801      &LkH$\alpha$\,198  &00$^h$11$^m$12.0$^s$  &+58\degr 54$'$01$''$    \\ 
HH\,802      &LkH$\alpha$\,198  &00$^h$11$^m$44.5$^s$  &+58\degr 42$'$39$''$    \\ 
HH\,803      &1548C27\,IRS\,1   &19$^h$42$^m$47.0$^s$  &+23\degr 22$'$19$''$    \\ 
HH\,804      &IRAS\,19395+2313  &19$^h$42$^m$10.4$^s$  &+23\degr 21$'$49$''$    \\ 
HH\,805      &IRAS\,19395+2313  &19$^h$41$^m$41.5$^s$  &+23\degr 20$'$34$''$    \\ 
HH\,806      &UNKNOWN           &19$^h$42$^m$03.4$^s$  &+23\degr 20$'$02$''$    \\ 
HH\,807      &UNKNOWN           &19$^h$42$^m$07.1$^s$  &+23\degr 19$'$54$''$    \\ 
HH\,808      &\LH               &22$^h$34$^m$35.6$^s$  &+40\degr 39$'$42$''$    \\ 
HH\,809      &\LH               &22$^h$34$^m$30.4$^s$  &+40\degr 39$'$01$''$    \\ 
HH\,810      &\LH               &22$^h$34$^m$21.2$^s$  &+40\degr 37$'$34$''$    \\ 
HH\,811      &\LH               &22$^h$34$^m$14.5$^s$  &+40\degr 36$'$48$''$    \\ 
HH\,812      &\LH               &22$^h$34$^m$11.6$^s$  &+40\degr 36$'$33$''$    \\ 
HH\,813      &\LH               &22$^h$34$^m$06.6$^s$  &+40\degr 36$'$18$''$    \\ 
HH\,814      &\LH               &22$^h$35$^m$01.4$^s$  &+40\degr 43$'$33$''$    \\ 
HH\,815      &\Lk               &21$^h$44$^m$29.9$^s$  &+66\degr 13$'$42$''$    \\ 
HH\,816      &\Lk               &21$^h$44$^m$26.4$^s$  &+66\degr 10$'$58$''$    \\ 
HH\,817      &\Lk               &21$^h$44$^m$13.3$^s$  &+66\degr 10$'$55$''$    \\ 
HH\,818      &\Lk               &21$^h$43$^m$57.7$^s$  &+66\degr 10$'$26$''$    \\ 
4H\,819      &\Lk               &21$^h$44$^m$01.0$^s$  &+66\degr 09$'$52$''$    \\ 
HH\,820      &\Lk               &21$^h$43$^m$47.9$^s$  &+66\degr 09$'$50$''$    \\ 
HH\,821      &\Lk               &21$^h$43$^m$43.4$^s$  &+66\degr 08$'$47$''$    \\ 
HH\,103\,A   &\Lk               &21$^h$42$^m$20.7$^s$  &+66\degr 03$'$31$''$    \\ 
HH\,822      &\Lk               &21$^h$41$^m$42.1$^s$  &+66\degr 01$'$45$''$    \\ 
HH\,823      &UNKNOWN           &21$^h$43$^m$27.9$^s$  &+66\degr 11$'$46$''$    \\ 
HH\,824      &IRAS\,21416+6556  &21$^h$42$^m$56.9$^s$  &+66\degr 09$'$10$''$    \\ 
HH\,825      &IRAS\,21416+6556  &21$^h$42$^m$39.2$^s$  &+66\degr 10$'$56$''$    \\ \hline
 
\end{tabular}
\caption{Positions of the new HH objects found in this survey and their probable sources.}
\label{HHPositions}
\end{table}

\section{Results for Individual Regions}
\subsection{LkH$\alpha$198 \& V376 Cas}

\indent LkH$\alpha$\,198 and its nearby companion, V376~Cas, are 
both Herbig Ae stars \citep{Herbig60} located in the small 
dark cloud L1265, at a distance of 600 pc \citep{Chavarria85}. 
An asymmetrical, bipolar molecular outflow in this region  was noted by 
\cite{Canto84}. \cite{Strom86} subsequently found the first optical outflow 
tracer, HH\,161, a bright HH object some 12\arcsec\ from LkH$\alpha$\,198 at a 
position angle (P.A.) of 100\degr$\!$. Further observations  
by \cite{Goodrich93} yielded another HH knot 81\arcsec\ away 
at a P.A. of 153\degr$\!$. This object was rediscovered by \cite{Aspin00} who 
refer to it as HH\,461. 

\begin{figure*}
\resizebox{\hsize}{!}{\includegraphics{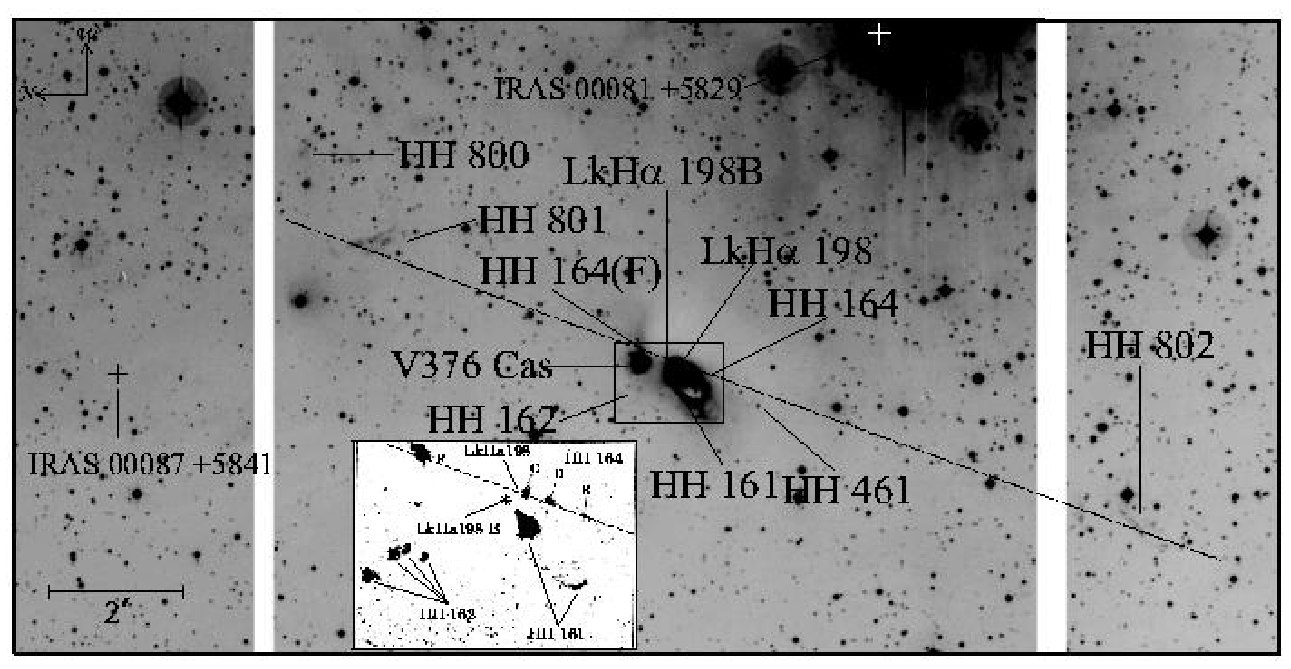}}
\caption{{\it LkH$\alpha$198 [SII] :\/} Mosaic of the entire outflow around 
LkH$\alpha$198, including the three newly discovered HH complexes with the 
main outflow axis delineated by a dotted line. For all images relating to 
LkH$\alpha$\,198 (Figs.\ \ref{LkHa198flow},\ \ref{LkHa198HH800-1} 
and\ \ref{LkHa198HH802}) north is to the left and west to the top but for all 
subsequent images, north is to the top unless specified otherwise. The continuum 
subtracted [SII] image of \cite{CRB95} is inset , showing the HH\,164 
knots (C, D, E and F), most of which are not seen in our [SII] image due to the 
presence of the reflection nebula. Here, the position of LkHa\,198 is 
indicated by a white cross slightly east of knot C. The P.A. of HH\,164 at 340\degr\ 
\citep{CRB95} is marked on the inset and it can be seen from the outflow 
axis on the main image that the P.A. of the extended outflow is also at 340\degr.
There are two known optically invisible IRAS sources in the region and their positions 
are marked. The white strips delineate gaps in the WFC CCD mosaic.}
\label{LkHa198flow}
\end{figure*}

The discovery, however, of LkH$\alpha$\,198\,B, a deeply embedded companion 
to LkH$\alpha$\,198, by \cite{Lagage93} raised the question of which of these
two stars is the primary outflow source in this region. In their study 
\cite{CRB95} (hereafter referred to as CRB) concluded there are two 
separate outflows with their origin in the vicinity of LkH$\alpha$\,198: 
one driven by LkH$\alpha$\,198 itself and the other by LkH$\alpha$\,198\,B. Their 
observations of HH\,161 revealed a tail pointing back towards 
LkH$\alpha$\,198\,B and they also discovered a suspected bow shock 39\arcsec\ 
southeast of this source. The bow shock (their knot B), HH\,161 and its tail 
are all aligned and appear to constitute a one-sided outflow from 
LkH$\alpha$\,198\,B. To date no counterflow has been seen. CRB also found a 
number of faint HH knots (HH\,164\,C, D and E -- see inset in 
Fig.\ \ref{LkHa198flow}) with the same P.A. from 
LkH$\alpha$\,198 as HH\,461. Thus HH\,164\,C, D and E and HH\,461 appears to 
delineate an outflow from LkH$\alpha$\,198. To the north they also discovered a 
faint knot (HH\,164 F) which may be a tracer of the counterflow from 
LkH$\alpha$\,198. Finally to the east of V376\,Cas a number of bright HH 
emission knots, HH\,162, were also found by CRB. These knots were seen again 
in our images, although their origin remains unclear. 

\begin{figure}
\resizebox{\hsize}{!}{\includegraphics{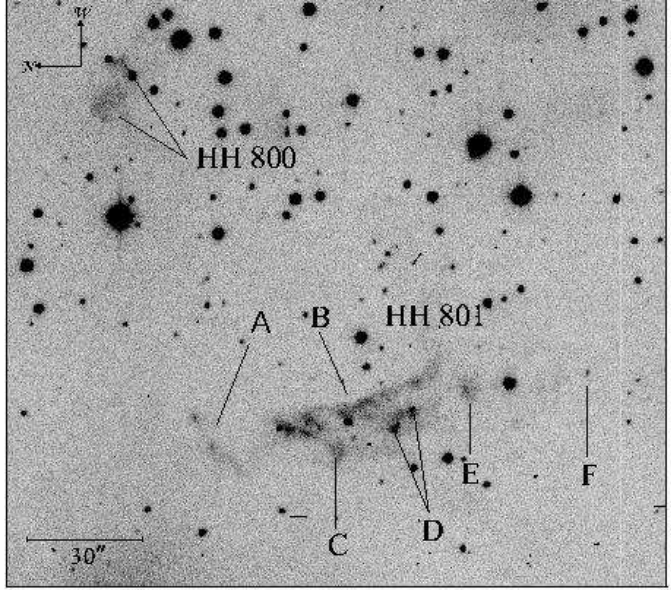}}
\caption{{\it LkH$\alpha$198 [SII] :\/} HH\,800 and HH\,801 showing various 
features referred to in the text. The knotty morphology of HH\,801 is 
clearly seen here.}
\label{LkHa198HH800-1}
\end{figure}

Two of the three HH complexes found here are extensions of the outflow from 
LkH$\alpha$\,198 discovered by CRB. HH\,801 has the same P.A. of 
340\degr\ with respect to LkH$\alpha$\,198 as HH\,164 F and is therefore 
almost certainly part of the same flow. It appears to be an asymmetrical 
bow shock of which we primarily see the western wing (B in 
Fig.\ \ref{LkHa198HH800-1}) some 45\arcsec\ in length. There are also a number 
of more easterly knots, C - F. Knot A may also be part of the western
wing. 

\begin{figure}
\resizebox{\hsize}{!}{\includegraphics{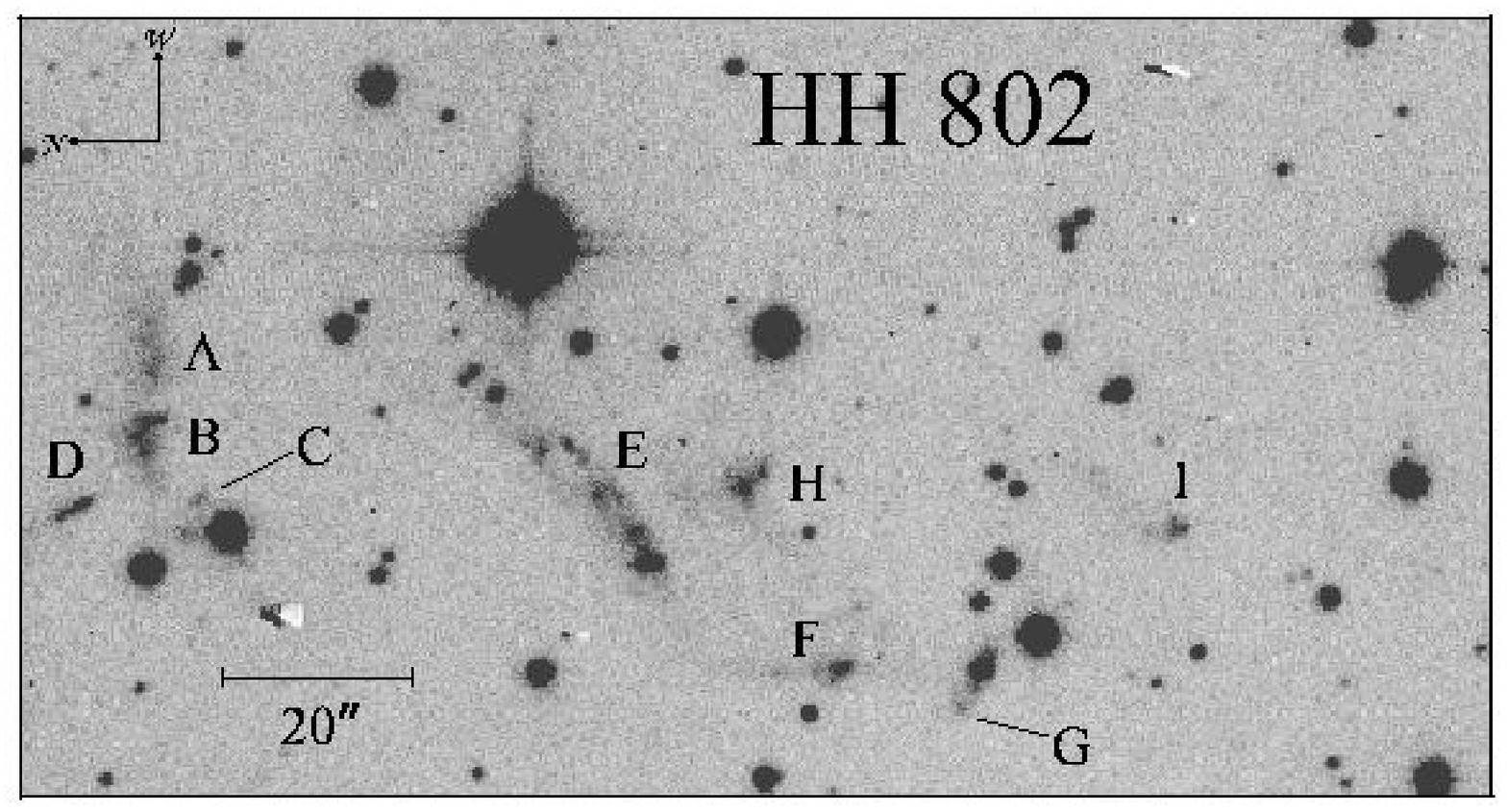}}
\caption{{\it LkH$\alpha$198 [SII] :\/} HH\,802, to the southeast of 
LkH$\alpha$\,198.}
\label{LkHa198HH802}
\end{figure}

In the diametrically opposite direction from HH\,801 through LkH$\alpha$\,198 
we find HH\,802. It consists of a number of features, A\,--\,I 
(Fig.\ \ref{LkHa198HH802}) which at first sight look 
somewhat chaotic. Feature E, however, like its counterpart in HH\,801, 
could be the western wing of an asymmetrical bow shock. 
The total length of HH\,802 is some 2$'$. It has a 
P.A. of 160\degr\  with respect to LkH$\alpha$\,198 and it is aligned with 
knots C, D and E (CRB) of 
HH\,164 and HH\,461 i.e.\ it is the counterflow of HH\,801 and HH\,164 Knot F. 
The furthest knot in HH\,802 (Knot I) is at a distance of 8\farcm16 (1.4\,pc) 
from LkH$\alpha$\,198, implying the total projected extent of 
the flow, from HH\,801 to HH\,802, is some 2.3\,pc.

HH\,800 to the northwest of HH\,801 is unlikely to be part of the 
HH\,801 -- HH\,802 outflow unless the outflow direction has changed abruptly. 
Although changes in flow direction have been observed in other parsec-scale 
outflows \citep{Reipurth01}, they tend to be more gradual. 
Moreover HH\,802 is even further from LkH$\alpha$\,198 than HH\,800 (at 
least in projection) and the flow associated with the former 
appears to have maintained a constant outflow direction. We should also 
add that HH\,800 is probably not part of the counterflow from 
LkH$\alpha$\,198\,B as its outflow is at a P.A. of 132\degr$\!\!$, 
while HH\,800 is at $\sim$330\degr\ with respect to this source. 
We cannot however exclude the possibility that the 
outflow axis may have swung through 18\degr. Finally either of the two optically 
invisible IRAS sources in the region (Fig.\ \ref{LkHa198flow})  
could be its driving source. IRAS completeness in this region is 
of the order of 5\Lsun$\!\!$. 
Proper motion studies would clearly help to identify its origin.

\subsection{1548C27}
\label{sec-results1548C27}

\indent The cometary-shaped reflection nebula 1548C27, and its associated 
\Ha emission line jet (HH\,165) were first noted by \cite{Craine81}. 
The optical jet, at a P.A. of 54\degr, was later confirmed
by \cite{Mundt84}. A low\,-\,mass, poorly collimated molecular outflow was 
also observed in the region by \cite{Dent92}.

\begin{figure}
\resizebox{\hsize}{!}{\includegraphics{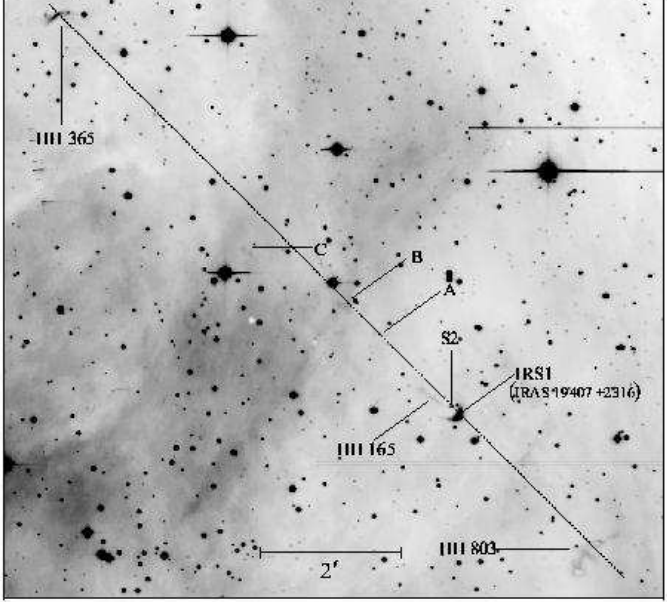}}
\caption{{\it 1548C27 \Ha :\/} Entire outflow around 1548C27\,IRS\,1. Note that
north is to the top and east to the left in this and subsequent images. An 
approximate outflow axis is marked here, at a P.A. of $\sim$ 45\degr, however 
the outflow appears to be precessing -- see text (section~\ref{sec-results1548C27}).}
\label{1548C27flow}
\end{figure}

Near-infrared photometry in the immediate vicinity of 1548C27 by 
\cite{Vilchez89} yielded two sources. One of these 
appears to be a foreground star but the other, IRS\,1, is located near the 
apex of the nebula, and they suggest this to be the driving source of 
HH\,165 (see Fig.\ \ref{1548C27flow}).
The IRAS PSC (point source catalogue) shows IRAS\,19407+2316 to be located 
close to, but not coincident with, IRS\,1. \cite{Vilchez89} however conclude 
that both near\,- and far\,-\,infrared sources are 
the same object, which, for convenience, we will refer to here as IRS\,1. 

The optical jet (HH\,165) is very narrow, with a width of 2$''$ - 3$''$ and 
its length is estimated to be $\sim$45$''$. There is a gap of $\sim$13$''$ 
between the source and the jet and two bright knots are visible at 23$''$ 
and 29$''$ from the source \citep{Mundt84}. The kinematic distance of 1548C27 
is 2.4 kpc \cite{Dent92} which implies a luminosity for IRS\,1 of approximately 
580\,L$_{\sun}$.

Another star S2, 10\arcsec\ northeast of 1548C27, was found by 
\cite{Scarrott91}. This star illuminates the nebula along with IRS\,1. 
\cite{Scarrott91} suggest HH\,165 curves towards this 
star, implying that it is the driving source, however we find no evidence 
in our images to support this idea.   

\begin{figure}
\resizebox{\hsize}{!}{\includegraphics{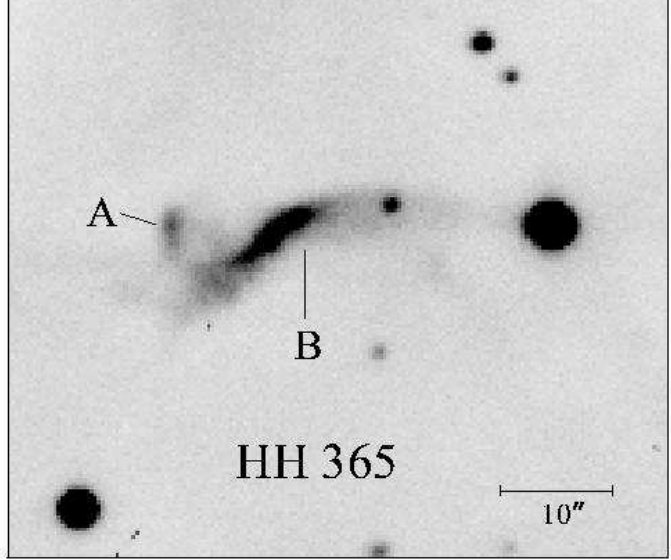}}
\caption{{\it 1548C27 \Ha :\/} HH\,365, to the northeast of 1548C27\,IRS\,1.}
\label{1548C27HH365}
\end{figure}

HH\,365 to the northeast of 1548C27 was briefly referred to by \cite{Alten97} 
as being bow-shaped and possibly associated with HH\,165. This object 
was mentioned in their paper but no image of it was included. Its
morphology is clearly seen here in Figs.\ \ref{1548C27flow} and\ 
\ref{1548C27HH365}.
HH\,365 is 8\farcm13 (5.7 pc) from IRS\,1 at a P.A. of 46\degr$\!$ i.e.\ 
close to that of the HH\,165 jet. From our 
images it appears to consist of two bright regions, Knot A and  
Feature B that extends to the northwest some 8$''$ (Fig.\ \ref{1548C27HH365}). 
The overall shape of Feature B is suggestive of an asymmetrical bow, the axis 
of which points roughly back towards 1548C27. 

\begin{figure}
\resizebox{\hsize}{!}{\includegraphics{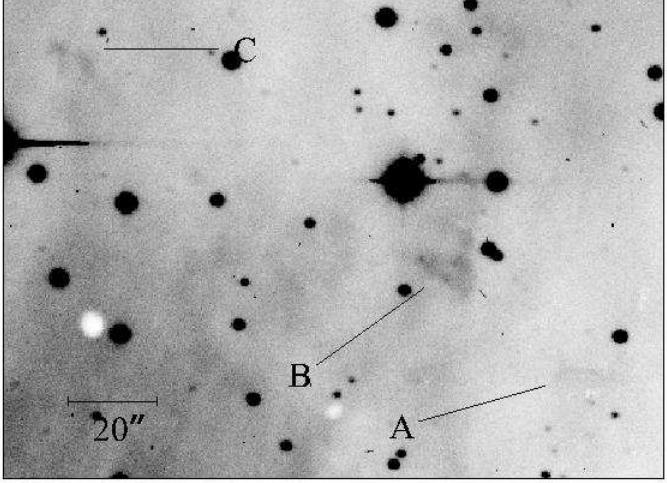}}
\caption{{\it 1548C27 \Ha :\/} Possible HH knots A, B and C to the northwest of 
HH\,165}
\label{1548C27ABC}
\end{figure}

Our survey also revealed a number of possible new HH objects 
(see Fig.\ \ref{1548C27flow}) although several are very faint. Moreover, it 
is unclear whether Knots A, B and C (Fig.\ \ref{1548C27ABC}), for example, to 
the northwest of HH\,165 are HH objects, as there is a lot of contaminating 
HII nebulosity in the region. However the fact that these lie on the path 
between HH\,165 and HH\,365 would suggest they might be. 
Further study is necessary, however, to determine their nature and for this 
reason we will desist from assigning them HH numbers. In any event it seems 
likely that HH\,165 and HH\,365 are part of the same outflow from IRS\,1 and 
that Knots A, B and C may also be part of this flow.

A counterjet from IRS\,1 was found recently in the near-infrared 
(Whelan, private communication). It lies along the same line as HH\,165 
at a P.A. of 234\degr\ and extends for at least 5\arcsec. A number of faint 
[FeII] emitting knots were also seen beyond the counterjet. Neither the 
counterjet nor any of these knots are observed in our optical images, 
presumably because of extinction.
 
\begin{figure}
\resizebox{\hsize}{!}{\includegraphics{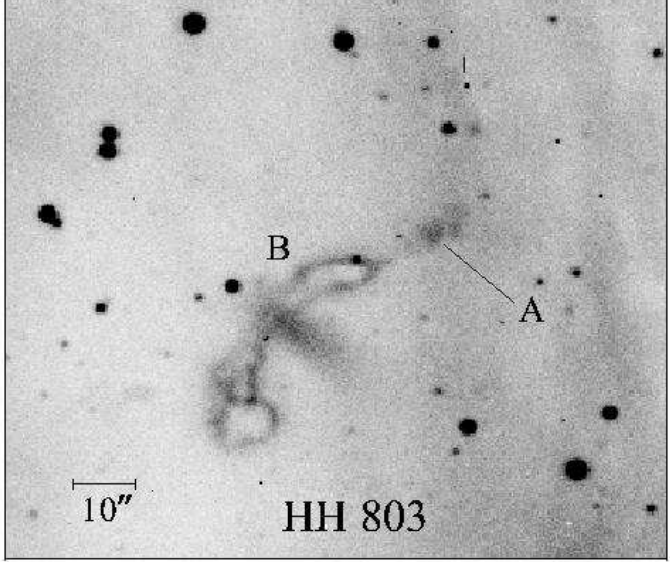}}
\caption{{\it 1548C27 \Ha :\/} The morphology of HH\,803 is clearly seen in 
optical images.}
\label{1548C27HH803}
\end{figure}

HH\,803 (Fig.\ \ref{1548C27HH803}), 2\farcm63 (1.85 pc) southeast of IRS\,1 at 
a P.A. of 223\degr$\!$, has a very interesting morphology. It appears to be 
bow-shaped but convex towards IRS\,1. It is 50\arcsec\ in width and 
contains a 13\arcsec\ ``jet\,-\,like'' feature bisecting the bow. The ``ring'' 
at the southern edge of the bow and the diffuse appearance of Knot A to the 
north add to the complexity of this object. Note that the jet\,-\,like feature
does not quite point back towards IRS\,1. We should also add that apart from 
IRS\,1, no other IRAS sources were found in the region strengthening the 
possibility that HH\,803 is driven by IRS\,1.

If we include Knots A, B and C as part of the HH\,165/HH\,365/HH\,803 outflow
then its overall appearance suggests that it may be slowly precessing with 
shifts in the outflow direction of at least 10\degr. The sense of 
precession (i.e. point-like symmetry through IRS\,1) to the northeast is also 
consistent with the position of HH\,803 to the southwest.

\subsection{The IRAS\,19395+2313 Region}

\begin{figure*}
\resizebox{\hsize}{!}{\includegraphics{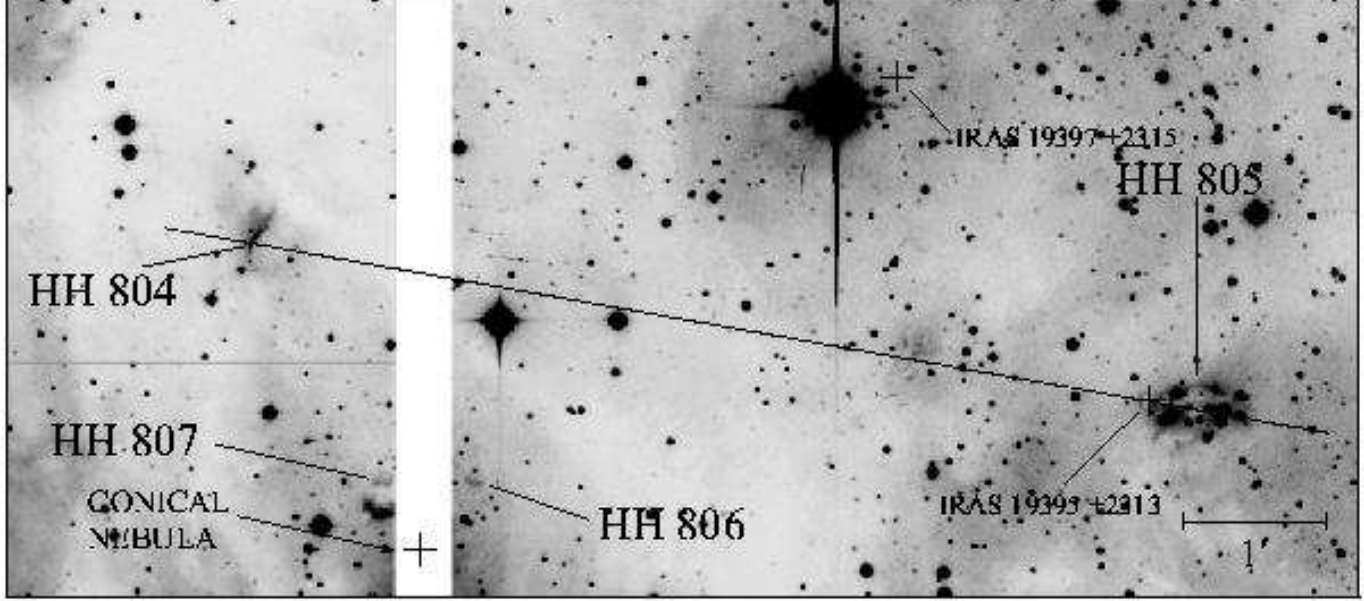}}
\caption{{\it IRAS\,19395+2313 \Ha :\/} The region around IRAS\,19395+2313, 
including all known additional IRAS sources in its vicinity. Note that IRAS 
completeness at this distance is $\geq$ 100\Lsun$\!\!$. Four newly 
discovered HH objects are marked. The position of the conical nebula, visible
on the Palomar Sky Survey Red (E) plate and which might be associated with the 
driving source for HH\,806, is indicated. The possible outflow from 
IRAS\,19395+2313 is marked with a dotted line at a P.A. of 80\degr.} 
\label{1548C27HH804-7}
\end{figure*}

A number of HH objects and possible sources were found in this region 
(see Fig.\ \ref{1548C27HH804-7}), which lies approximately 18$'$ west of 
1548C27 and is also in the vicinity of the young open cluster NGC\,6823. 
It is highly unlikely that any of these newly-discovered 
HH objects are part of the 1548C27\,IRS\,1 outflow  although we 
will assume they are at the same distance, i.e.\ 2.4 kpc. 

Two of the newly discovered HH objects are found close 
together - HH\,806 is 30\arcsec\ west of HH\,807 (Fig.\ \ref{1548C27HH804-7}).
The region between them coincides with a gap in the CCD mosaic 
although a cursory inspection of the Palomar Sky Survey Red (E) plate shows 
there is a star in the ``gap'', ALS\,10422 or IRAS\,19399+2312 (at 
19$^h$42$^m$05.5$^s$ +23\degr 18$'$59$''$, J2000) which, at first sight, 
might be a possible HH driving source. This object is however 
an AGB star \citep{Parthasarathy00} so we can disregard it. 
We also see on the Palomar Sky Survey red (E) plate, a 
conical nebula $\sim$36\arcsec\ southeast of HH\,806 
(at 19$^h$42$^m$02.12$^s$ +23\degr 19$'$30$''$, J2000) which may be 
associated with the driving source of this object. In fact HH\,806 lies along 
the major axis of this conical nebula. Although this nebula is not seen in 
Fig.\ \ref{1548C27HH804-7}, its position is marked. There is no obvious 
driving source for HH\,807.  

\begin{figure}
\resizebox{\hsize}{!}{\includegraphics{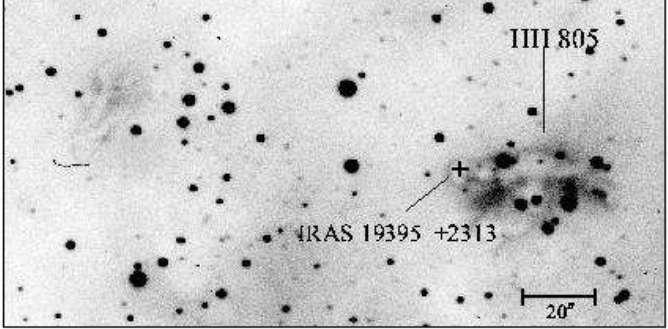}}
\caption{{\it IRAS\,19395+2313 \Ha :\/} HH\,805 and its candidate driving 
source IRAS\,19395+2313.}
\label{1548C27HH805}
\end{figure}

HH\,805 has an interesting morphology (Fig.\ \ref{1548C27HH805}) and 
IRAS\,19395+2313 seems an obvious candidate to be its driving source given its 
position. We have estimated the luminosity of IRAS\,19395+2313 to be $\sim$ 320 
\Lsun$\!\!$. The angular extent of HH\,805 is approximately 45\arcsec\ (0.5 pc) and it 
can be seen from Fig.\ \ref{1548C27HH805} that this outflow is poorly 
collimated. Morphologically it appears to have a knotty ring-like structure and
is reminiscent of the HH complex associated with V380\,Ori \citep{Corcoran95}. 
HH\,804 (Fig.\ \ref{1548C27HH804-7}) may be part of the counterflow from 
IRAS\,19395+2313 although we emphasise that this association is {\em highly} 
uncertain.  It is at a P.A. of 
80\degr\ with respect to the latter. HH\,804 is at a distance of 6\farcm3 
from IRAS\,19395+2313, implying the total projected extent of the flow, 
assuming HH\,804 is part of it, is $\sim$5 pc.

\subsection{LkH$\alpha$ 233}

\indent \LH is an A5e\,-\,type pre\,-\,main sequence star \citep{Corcoran97} 
at a distance of 880 pc and is associated with a bipolar nebula that is 
approximately 0.1 pc in size \citep{Calvet78,Staude93}. The nebula has a 
distinct X\,-\,like morphology with bright reflection 
limbs at 50\degr$\!$\,/\,90\degr and 230\degr$\!$\,/\,270\degr$\!$.

\begin{figure}
\resizebox{\hsize}{!}{\includegraphics{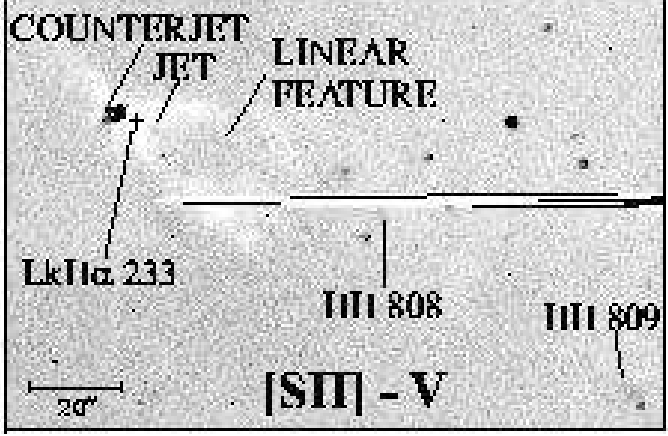}}
\caption{{\it LkH$\alpha$ 233 [SII]:\/} Continuum subtracted image of the 
centre of the LkH$\alpha$ 233 nebula. The jet is optically visible here, along 
with a number of other emission features close to the source including HH\,808 
and HH\,809.}
\label{LkHa233center}
\end{figure}

The discovery by \cite{Leinert93} of a light scattering ``halo'' 
$\sim$1$''$ in size around \LH led them to suggest that the star is 
highly embedded and optically visible largely through scattered light.
\cite{Corcoran98} discovered a bipolar jet and counterjet (HH\,398)
spectroscopically at P.A.s of approximately 245\degr\ and 
65\degr\ that bisect the X\,-\,shaped nebula. 
In their spectrograms, the redshifted counterjet is seen to begin $0''\!\!.7$ 
from the centre of the stellar continuum, whereas the blueshifted jet can be 
traced right back to the continuum peak. The jet, and counterjet, 
can be seen in our continuum subtracted image (Fig.\ \ref{LkHa233center}) 
along with a number of HH objects along the outflow direction. 

Evidence for the presence of a large ``polarisation disk'' centred on 
\LH with a radius of about 15000 AU was found by \cite{Aspin85}. The 
position angle of this disk is about 155\degr$\!$, which 
places it perpendicular to the observed optical outflow. The existence of 
this ``polarisation disk'' combined with the fact that the counterjet is not 
seen spectroscopically close to the star leads to the conclusion that 
there is a circumstellar disk present that obscures the beginning of the 
receding flow.  

Close to \LH the [SII] emission can be resolved into two velocity 
components \citep{Corcoran98}. The high velocity component is identified 
with the optical jet while the low velocity component, which
extends to less than 2$''$ from the star, may be modelled as a 
rotationally broadened disk wind (see Kwan \& Tademaru, 1995).

\begin{figure*}
\resizebox{\hsize}{!}{\includegraphics{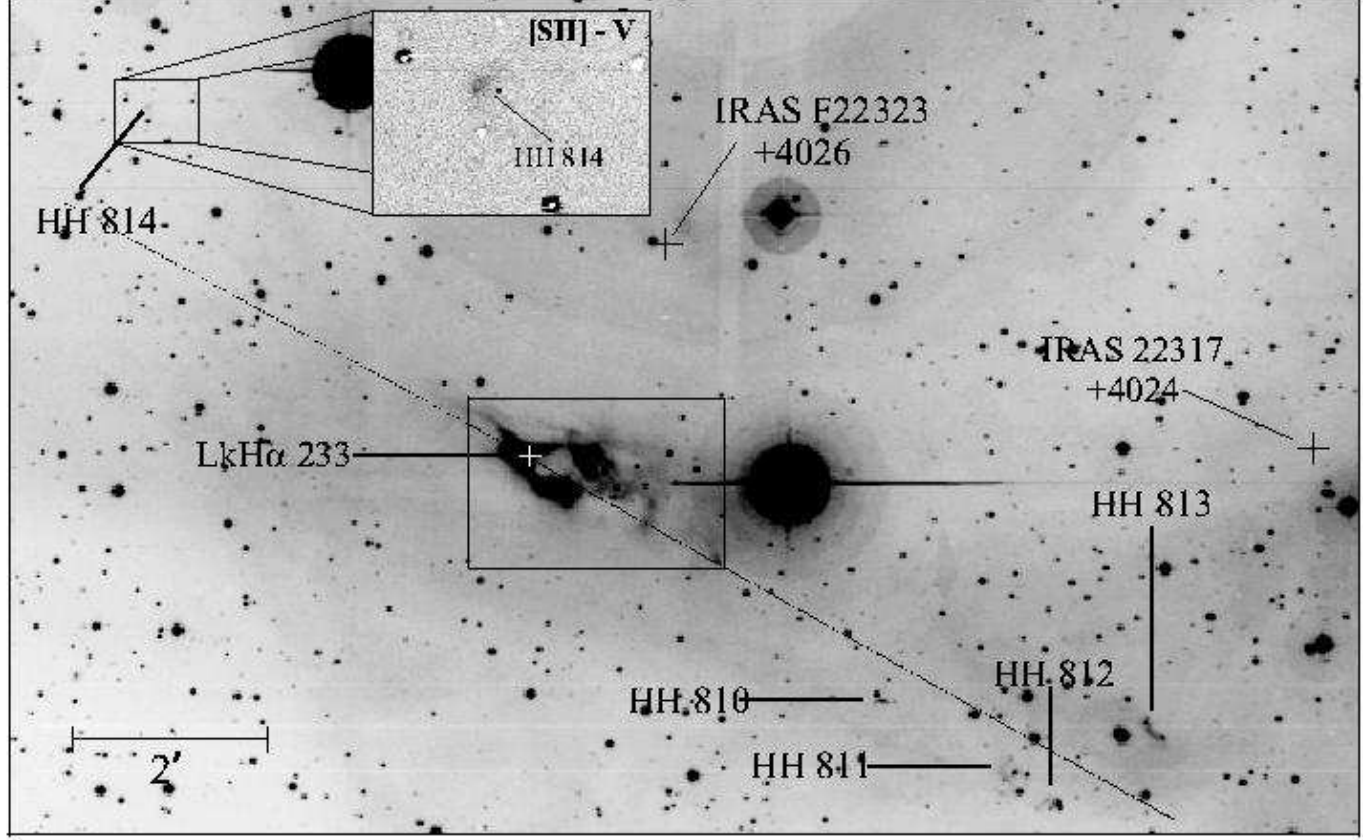}}
\caption{{\it LkH$\alpha$ 233 [SII] :\/} The entire outflow around \LH 
including all known optically invisible IRAS sources in the region. The major 
axis of the outflow at 62\degr\ through the source is indicated by a straight
line. The continuum subtracted ([SII]-V) image in the top right corner
shows HH\,814 more clearly. The area around LkH$\alpha$ 233 is marked by a 
box and is seen in more detail in Fig.\ 12.} 
\label{LkHa233flow}
\end{figure*}

This survey revealed a number of previously unknown HH objects in the vicinity 
of \LH$\!\!$. It is possible that not all of these objects can be 
attributed to \LH and the positions of a nearby 
IRAS source, IRAS\,22317+4024, complicates our analysis 
(see Fig.\ \ref{LkHa233flow}). Candidate driving sources for all the 
new HH objects are suggested here. 

Continuum subtracted images of the nebula surrounding LkH$\alpha$ 233 reveal a 
number of HH features which are not seen, at least so clearly, in the [SII] 
image alone. Fig.\ \ref{LkHa233center} 
shows the first optical images of an $\sim$7$''$ jet emerging from the 
\LH nebula at a P.A. of $\sim$248\degr, which is relatively close to the 
P.A. of the jet as spectroscopically determined by \cite{Corcoran98}. A 
possible counterjet to the northeast of \LH is also seen in this image. 
But it is difficult to determine whether this is actually a counterjet or 
simply a residual of the continuum subtraction process. 
There are two other emission knots to the southwest of the 
source. The first of these, HH\,808, is situated close to a diffraction 
spike from a bright star to the west of \LH$\!\!$. HH\,808 is 1\farcm05 
from the source at a P.A. of 250\degr. The second knot, 
HH\,809, is 2\farcm2 from the source at a P.A. of 241\degr. 
Two other objects were seen to the northeast of \LH at a distance of 
$\sim$2$'$, the first has a P.A. of 63\degr\ 
and the second is at 65\degr. These objects are outside the area shown in 
Fig.\ \ref{LkHa233center} and it is unclear at present whether these are 
HH objects.

Also to the  northeast of \LH$\!\!$, we discovered HH\,814 
(Fig.\ \ref{LkHa233flow}) at a distance of $5'\!\!.18$ (1.3 pc). The morphology 
of this object may be studied more clearly in the continuum subtracted 
image inset in Fig.\ \ref{LkHa233flow}. HH\,814 appears to be a 
broad bow shock facing back towards \LH$\!\!\!\!$ at a P.A. of 47\degr\ with 
respect to \LH . \cite{Corcoran98} determined a 
P.A. of approximately 65\degr\ for the counterjet, suggesting that if 
HH\,814 is part of the same flow, its direction has changed by 
$\sim$20\degr$\!$. Note however that the P.A. determined by \cite{Corcoran98}
is very crude as it was deduced by slit sampling at various P.A.'s. 

\begin{figure*}
\resizebox{\hsize}{!}{\includegraphics{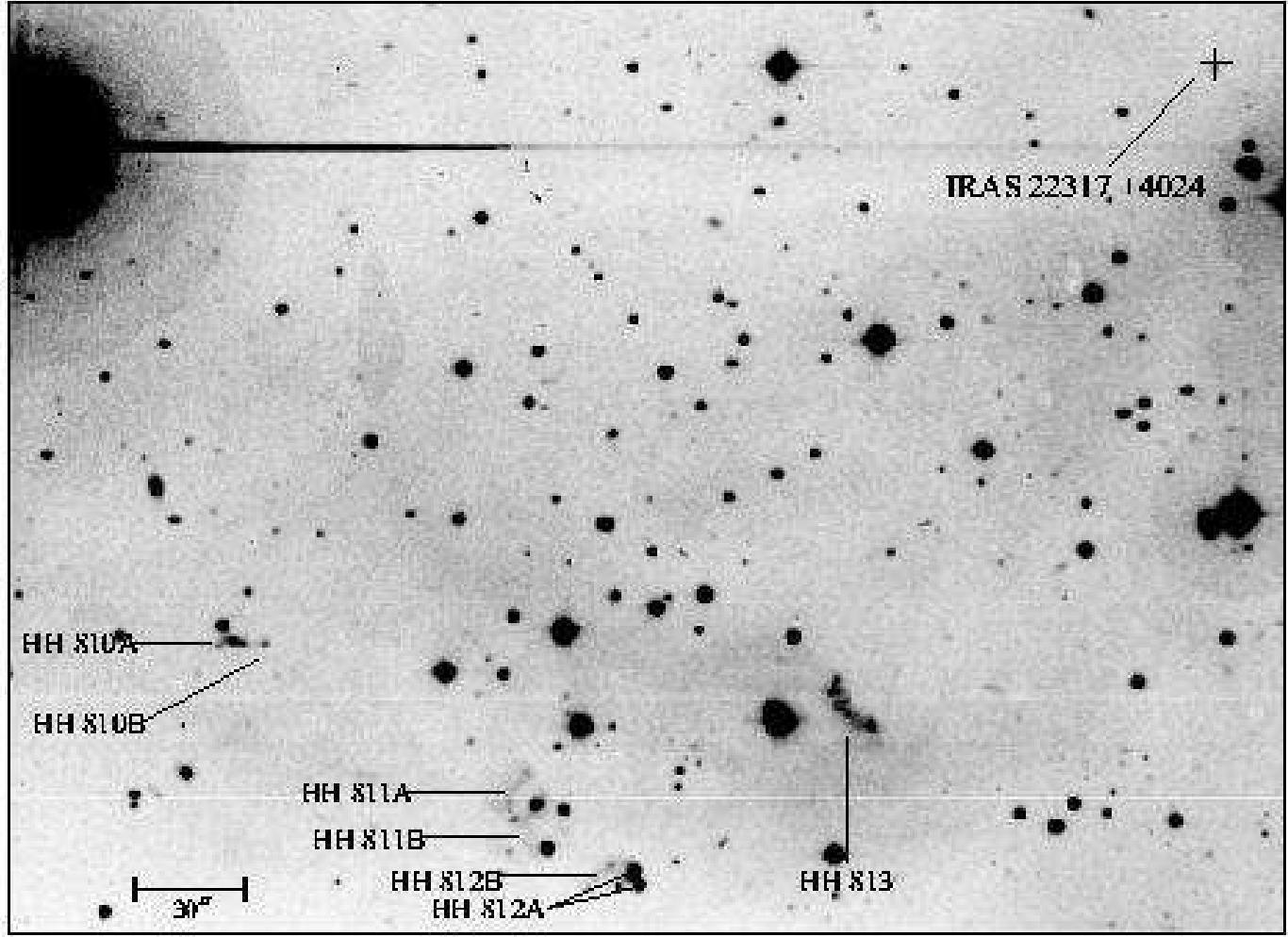}}
\caption{{\it LkH$\alpha$ 233 [SII]:\/} HH\,810\,--\,HH\,813, 
to the southwest of \LH$\!\!$.}
\label{LkHa233HH813-11}
\end{figure*}

A number of other objects were discovered to the southwest of \LH (see 
Fig.\ \ref{LkHa233HH813-11}). HH\,810, HH\,811 and HH\,812, are at 
4\farcm5 (1.2 pc), 6$'$ (1.5 pc) and 6\farcm5 (1.7 pc) respectively 
from \LH$\!\!$, all at a P.A. of 236\degr. HH\,813, at a distance
of 7$'$ (1.8 pc) has a P.A. of 247\degr\  with respect to 
\LH$\!\!$. Considering the possibility that outflows from higher mass 
stars may not be as collimated as those from low mass stars, 
HH\,810\,-\,HH\,814 could be the optical tracers of the edges of a 
moderately collimated flow driven by \LH$\!\!$. The axis of this outflow 
with respect to \LH is $\sim$62\degr$\!$ (marked on Fig.\ 
\ref{LkHa233flow}), which agrees well with the estimate of the P.A. of the 
jet closer to the source. It is also possible that HH\,813 is a bow shock 
facing back towards IRAS\,22317+4024 (Fig.\ \ref{LkHa233HH813-11}). Proper 
motion studies could conclusively determine whether this is the case.

\subsection{The NGC\,7129 region}

\indent A large number of YSOs are known in the NGC\,7129 cluster 
which lies at a distance of 1.25 kpc \citep{Shevchenko89}. Aside 
from optically visible young stars such as \Lk$\!\!$, there are many 
embedded ones. For example, a 160\,$\mu$m survey by \cite{Bechis78} 
revealed two far infrared sources and although one of them is spatially 
coincident with \Lk (FIRS\,1), the other is ~3$'$ further south (FIRS\,2)
and optically invisible. Additional infrared \citep{Weintraub94,Cabrit97} and
sub\,-\,millimetre sources \citep{Fuente01} are also known. 

Associated with these sources one finds the usual host of phenomena 
typical of star formation including reflection nebulae \citep{Bertout87}, 
molecular outflows \citep{Edwards83,Bertout87,Mitchell94}, HH objects
\citep{Ray87,Miranda93}, HH jets \citep{Ray87,Ray90,Cabrit97} and shocked 
H$_2$ flows \citep{Eisloffel00}.

Looking at the distribution of previously known HH objects in the region, 
and the new ones discovered here (see below), one gets the  
impression (see Figs.\ \ref{LkHa234flow} and \ref{LkHa234centre}) 
that the primary outflow axis, or axes, is roughly in a northeast to 
southwest direction and centred on the cluster core. Caution however is 
necessary. Proper motions studies have shown that some 
HH objects like GGD\,32 and HH\,103 are not moving to the southwest, as 
one might suspect, but instead to the west \citep{Ray90}. Moreover others, 
like GGD\,34 \citep{Gomez99} and possibly GGD\,33 \citep{Cohen83,Goodrich86}, 
have their own sources outside the cluster core. 

\begin{figure}
\resizebox{\hsize}{!}{\includegraphics{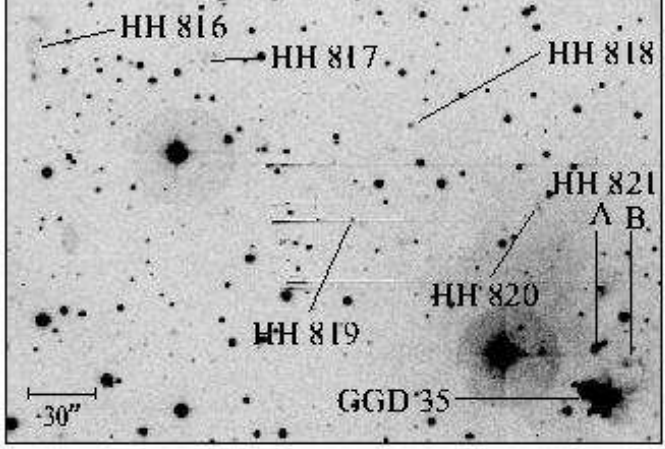}}
\caption{{\it LkH$\alpha$ 234 [SII]:\/}  HH\,816 -- HH\,821 (A and B), to 
the northeast of \Lk$\!\!$.}
\label{LkHa234NE}
\end{figure}

The HH\,815 complex (Fig.\ \ref{LkHa234NE2}) is over 1$'$ in size and is at a 
distance of 11$'$ (4 pc) from \Lk$\!\!$. The three emission regions in 
HH\,815 (A, B and C) appear to form the edges of a large bow shock that is 
concave towards the cluster core. HH\,816 (Fig.\ \ref{LkHa234NE}) may be 
another bow trailed by a series of faint HH knots i.e.\ HH\,817\,-\,HH\,820. 
HH\,821, $\sim$15$''$ north of GGD\,35, is aligned with HH\,819 and HH\,816 
and so could form part of the same flow. We note also that HH\,105, is 
on the same axis from the cluster core and that no source for HH\,105, at 
least in its immediate vicinity, is known.

\begin{figure}
\resizebox{\hsize}{!}{\includegraphics{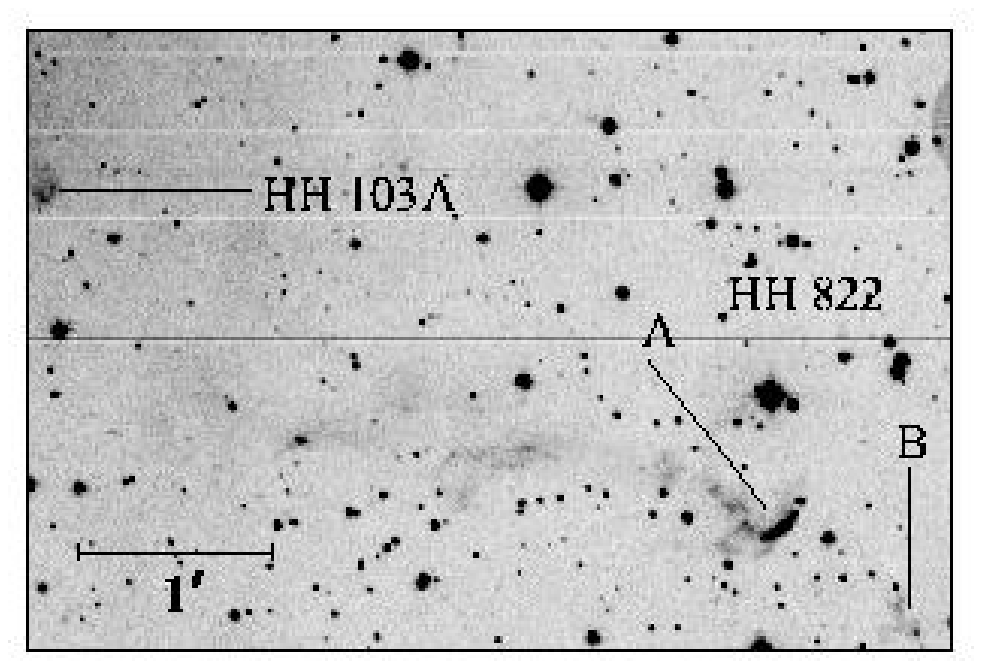}}
\caption{{\it LkH$\alpha$ 234 [SII] :\/} HH\,103\,A and HH\,822 to the 
southwest of \Lk$\!\!$.}
\label{LkHa234SW}
\end{figure}

The most distant HH object to the southwest of the cluster discovered by us is 
HH\,822 (Fig.\ \ref{LkHa234SW}). It is at a P.A. of 238\degr\ with respect to 
\Lk and the morphology of its brightest component (HH\,822~A), i.e.\ a bow 
concave towards the cluster, suggests it is part of an outflow that originated 
there. An additional knot close to HH\,103 was also found and we shall refer 
to it as HH\,103\,A (Fig.\ \ref{LkHa234flow}). Given its location, it seems 
likely it is part of the same flow that drives HH\,822. Moreover
faint emission can be seen linking HH\,103\,A to HH\,822\,A reinforcing 
this conclusion. Finally we add that HH\,822\,B is at a distance of 
10\farcm7 (3.9 pc) from \Lk$\!\!$.

The large number of sources in this region makes it extremely 
difficult, without detailed kinematic studies, to determine the origin 
of individual HH complexes. As previously mentioned there are a number 
of low mass YSOs, like the one that drives GGD\,34, present and this 
complicates our analysis. That said, it seems likely, purely on 
morphological grounds as well as location, that many of the newly
discovered HH objects are driven by a source(s) in the cluster core. 
Because of their large distance from the core, however, it may prove 
impossible, even with good proper motion data, to determine their precise
origin. Several possible candidates exist including \Lk$\!\!$, IRS\,6 and 
FIRS\,1\,-\,MM1. If we draw an imaginary axis through the core 
at a P.A.\ of 60\degr$\!$\,/\,240\degr\ it roughly  
delineates the region where most of the new HH objects, HH\,816 to HH\,822, 
are located.  
Assuming we are dealing with one outflow here, that originates in the core, 
then its overall angular size is 21\farcm8 i.e.\
$\sim$8 pc in projected length. 

There are three other HH objects which are not situated along the major axis 
marked in Fig.\ \ref{LkHa234flow}. HH\,824 and HH\,825 are located on either 
side of IRAS\,21416+6556 at 143\degr\ / 323\degr\ with respect to this source, 
suggesting a possible bipolar outflow driven by IRAS\,21416+6556.
These objects are 2\arcmin\ and 51\arcsec\ respectively from IRAS\,21416+6556.
The driving source of HH\,823 is unclear as is its association with any of the 
other outflows in this region.

\begin{table*}
\begin{tabular}{llllllllll}
\hline \hline
Source &Distance &Ref.  &L$_{bol}$  &Ref. &Outflow  &Ref. &Associated  &\CS       &\tf \\
        &(pc)     &      &/\Lsun   &     &length (pc) &  &Cloud       &(pc)       &(\degr)      \\ \hline
\1     &600      &2    &$\geq$ 160  &2   &2.3    &1    &L1265          &2     &2.8          \\  
\2     &2400     &3    &580         &3   &7.5    &1    &NGC\,6823      &33        &11.6         \\  
\3     &2400     &3    &320         &1   &5      &1    &NGC\,6823      &33       &30.2         \\  
\LH  &880      &4    &$\geq$ 121  &4   &3.1    &1    &ANON           &3.2         &6.3          \\  
\Lk    &1250     &5    &1200        &6   &8      &1    &NGC\,7129      &5.1       &6.2          \\  
\4     &1700     &7    &1700        &8   &5.3    &8    &L291           &15        &1            \\  
\5     &750      &9   &120         &9   &2.4    &10   &L1165          &1.8        &4.04         \\  
\9     &500      &11   &100         &12  &2.6    &10,13 &L1617         &28        &6.7          \\ 
\8     &460      &10   &25          &10  &7.7    &10   &L1617          &28        &7.4          \\ 
\6     &460      &14   &50          &15   &5.9    &16   &L1641          &25       &10.3         \\ 
\7     &460      &14   &28          &17  &3      &18   &L1641          &25        &1.5          \\  \hline
\end{tabular}
\caption{Parameters of newly discovered, and some previously known, parsec\,-\,scale outflows from low\,- and 
intermediate\,-\,mass YSOs.
\newline$^a$ : This is the diameter of the cloud where diameter = [major axis + minor axis]/2
\newline$^b$ : $\theta$$_{flow}$ is calculated by taking the width of the most distant shock in both the 
outflow and counterflow and dividing by the projected distance from the source. The mean value of $\theta$$_{flow}$ 
for the outflow and counterflow is given here. 
\newline$^c$ : The source of HH\,80\,/\,HH\,81.
\newline$^d$ : The source of HH\,111.
\newline$^e$ : These YSOs are low\,-\,mass sources and are included here for comparison purposes only.
\newline References : 1. this paper; 2. \cite{Chavarria85}; 3. \cite{Dent92}; 4. \cite{Calvet78}; 5. \cite{Shevchenko89};
6. \cite{Harvey84}; 7. \cite{Rodriguez80}; 8. \cite{Marti93}; 9. \cite{Schwartz91}; 10. \cite{Reipurth97}; 11. 
\cite{Cohen81}; 12. \cite{Mundt94}; 13. \cite{Gomez97}; 14. \cite{Hester98}; 15. \cite{Harvey86}; 16 \cite{Ogura95}; 
17. \cite{Reipurth93}; 18. \cite{Bally94}}
\label{parameters}
\end{table*}

\section{Discussion}
\subsection{Overall lengths}

It has been stated, albeit somewhat tongue-in-cheek, that the apparent length 
of optical outflows from YSOs used to be a function of CCD 
chip size! Early chips were small and sampled only a small angular patch of 
the sky. This, in combination with the episodic nature of the flows themselves, 
conspired to suggest outflow lengths measured in tens of thousands of AU rather
than parsecs. In a number of cases however CCD mosaicing \citep{Ray87} did 
hint that some flows were at least in the parsec league. Large format CCDs in 
focal\,-\,plane mosaics can cover fields of view larger than 30\arcmin\ which, 
at a distance of 1kpc, corresponds to more than 8.7pc. Flows can therefore be 
detected to beyond the peripheries of their parent cloud.

Table\ \ref{parameters} lists parameters such as distance, source luminosity, 
outflow length, cloud size and degree of collimation for both the 
intermediate\,-\,mass sources discussed here and several well-studied outflows 
powered by both intermediate\,- and low\,-\,mass YSOs. In some regions, the 
projected lengths of the outflows are similar to the sizes of the clouds from
which they emerge. This correlation of length scales is to be expected and there 
should also be a tendency for shocks to be seen near the cloud edges where 
extinction by dust is minimal. 

If we assume an average tangential velocity for outflows of 100 kms$^{-1}$, 
in 10$^5$~yrs material originally at the source will be transported $\sim$10 pc 
i.e.\ typically to the edge of the clouds we are studying. Using the 
evolutionary tracks of \cite{Palla93}, we see that such a period corresponds to 
about 10\% of the time an average Herbig Ae/Be star spends in the pre-main
sequence phase. It follows that only in the case of the more massive YSOs, 
i.e.\ those with the shortest evolutionary timescales, do HH outflows
represent a ``fossil record'' of activity over the totality of the outflow 
phase. 
  
\subsection{Morphology}

HH outflows are episodic: they are clearly not continuous phenomena. 
For the most part, the shocks we see are generated by supersonic jet material 
ramming into previously ejected slower gas. This process produces a series 
of ``working surfaces'', radiative shock systems that fade with time and 
therefore with distance from their source. Only the strongest shocks survive
to produce dramatic, often chaotic, structures on parsec scales. It has 
even been suggested that the FU Orionis phenomena may signal the dramatic 
change in output needed at the source to produce such a shock 
\citep{Reipurth89}. Thus the spacing between HH objects increases with 
distance from the source at least amongst low mass YSOs \citep{Reipurth01}. 
Such a trend is also visible here amongst the flows from intermediate mass 
YSOs such as LkH$\alpha$\,198 and 1548C27. Another phenomenon that occurs 
with HH outflows from low mass YSOs is that the shock structures become 
larger and apparently more chaotic with distance. Again this is something 
which is replicated in their intermediate mass YSO counterparts.  

A phenomenon that is found in parsec-scale outflows from lower mass YSOs is 
that the flow often exhibits ``S'' or ``C'' shape symmetry \citep{Reipurth89} 
possibly as a result of jet precession or source motion through the parent 
cloud respectively. In our small sample, we do not find any clear-cut 
examples of either although, as previously mentioned, the parsec-scale 
outflow from 1548C27\,IRS\,1 may be ``S-shaped''. 

\subsection{Collimation}

As alluded to in the Introduction, one of the most striking differences 
between parsec-scale outflows from low and massive YSOs is the relative 
lack of collimation seen in the latter \citep{Shepherd97,Hunter97,Shepherd98}.
This point is well illustrated by the archetypal example of an outflow from 
a high luminosity YSO: the Orion IRC2 flow \citep{Allen93}. Its opening angle 
is $\sim$ 90\degr\ \citep{Burton94} which is a sharp contrast to the 
angles (typically a few degress) seen in outflows from low\,-\,mass YSOs.  

Table\ \ref{parameters} list ``final'' opening angles for our small sample.  
Values are determined by dividing the width of the most distant HH 
object by its projected source separation. Note that as we are using projected 
separations the quoted values must be upper limits, however our observations 
only reveal the brightest portions of the shocks and fainter outer parts may be 
missed, so the opening angles may actually be greater than what is optically 
observed. Observed opening angles range from 0.9\degr\ to 12\degr\ suggesting 
a degree of collimation comparable to that seen in outflows from low\,-\,mass YSOs.   
This also suggests that the transition from well\,-\,collimated outflows to 
poorly\,-\,collimated outflows occurs at higher masses than the sources observed 
here.

\subsection{The Frequency of Blow\,-\,outs}
The true size of an outflow, in comparison to that of its parent cloud, is an 
important factor in determining whether the outflow's energy and momentum 
is transported into the ISM or remains within the cloud itself. A cursory 
examination of our data shows a clear tendency for HH objects to lie close to 
the edges of the parent cloud or at least close to the edges of clumps. As 
already mentioned, such an effect is to be expected considering the low 
extinction near cloud peripheries and the very low ISM densities beyond the 
cloud boundaries. 

More importantly, it is clear that actual outflow timescales are very long 
in comparison to apparent dynamical ones. Here dynamical timescales are 
derived by dividing the projected length of an outflow by its estimated 
tangential velocity. This, and the observation of HH objects near cloud 
boundaries, immediately suggests that most, if not all, of the flows studied 
here have blown-out of their parent cloud.  

\subsection{Are outflows a source of cloud turbulence?}
There is plenty of evidence to suggest molecular clouds are turbulent 
(Ward\,-\,Thompson 2002 and references therein) and that the pressure
generated by this turbulence is sufficient to prevent clouds from collapsing
under their own gravity. It has been shown however that cloud
turbulence, even in the presence of magnetic fields, decays too quickly 
compared to typical cloud lifetimes \citep{Stone98} and must therefore be 
somehow replenished. Could outflows be the primary source of turbulence in 
molecular clouds? 

The parsec-scale HH outflows imaged here appear largely well-collimated and 
therefore one might think they could affect only a narrow cone of 
ambient material. One has to remember however that these flows are supersonic 
and that we are viewing only the 
most highly collimated outflow component. The same flows ``imaged'' in 
the CO J = 1$\rightarrow$0 line would normally appear much less collimated, 
especially at low velocities. That these flows affect cloud structure on parsec
scales is evident from features such as the large CO ``cavity'' in NGC 7129. No 
doubt they also affect cloud dynamics. \cite{Arce01} for example has found that 
molecular outflows associated with parsec-scale HH flows can 
possess kinetic energies comparable to the turbulent and gravitational binding 
energies of their parent clouds.  This proves however only that they are a 
potentially important source of turbulence. Unfortunately we do not understand
the coupling between outflows and their ambient medium well enough to be sure.
Numerical simulations \citep{Downes03} are helping to address this problem but 
further studies are required.

\section{Conclusions}

We have investigated the occurrence of parsec-scale outflows from 
intermediate\,-\,mass YSOs. As is the case with lower mass YSOs, 
such flows appear to be common and we report the discovery of 
four here with well defined sources. These include LkH$\alpha$\,198, 
1548C27\,IRS\,1, \LH\, and IRAS\,19395+2313.  All of these YSOs, with 
the exception of the last, were previously known to have small-scale 
outflow activity. 

The region surrounding the Herbig Ae star \Lk is cluttered with outflows and 
candidate sources. Twelve new HH objects are reported on here, with many of 
them lying along a preferential axis centred near \Lk and orientated in a 
northeast-southwest direction. Their morphology suggests that at least 
some are part of a large-scale flow centred on the core of the 
NGC\,7129 cluster.

Parsec\,-\,scale outflows from intermediate\,-\,mass YSOs show a number of 
similarities to those from their low\,-\,mass counterparts. In particular :\\
- their lengths and degree of collimation appear comparable, \\
- they share the same morphological trends such as decreasing 
frequency, increasing dimension and increasing complexity of HH emission 
with distance from the source.\\

The lengths of these large-scale outflows are usually comparable to the 
clump size of their associated clouds. As their expected 
lifetimes are much larger than the apparent dynamical timescales, this 
suggests that many have ``blown-out'' of the cloud complex.

Finally, it is evident that the transition from highly collimated jet\,-\,like 
flows to poorly collimated wide\,-\,angle outflows such as OMC\,1 must lie at 
higher masses and luminosities than the sources studied here.

\acknowledgement{
FMcG and TPR acknowledge support from Enterprise Ireland. JBs research was 
supported by NSF grant AST-9819820, NASA grant ANG5-8108 (LTSA), 
and NASA grant NCC2-1052 (CU Center for Astrobiology).}

\clearpage

\clearpage

\clearpage

\clearpage 

\clearpage

\clearpage

\clearpage

\clearpage

\clearpage

\clearpage

\clearpage 

\clearpage

\begin{figure*}
\resizebox{\hsize}{!}{\includegraphics{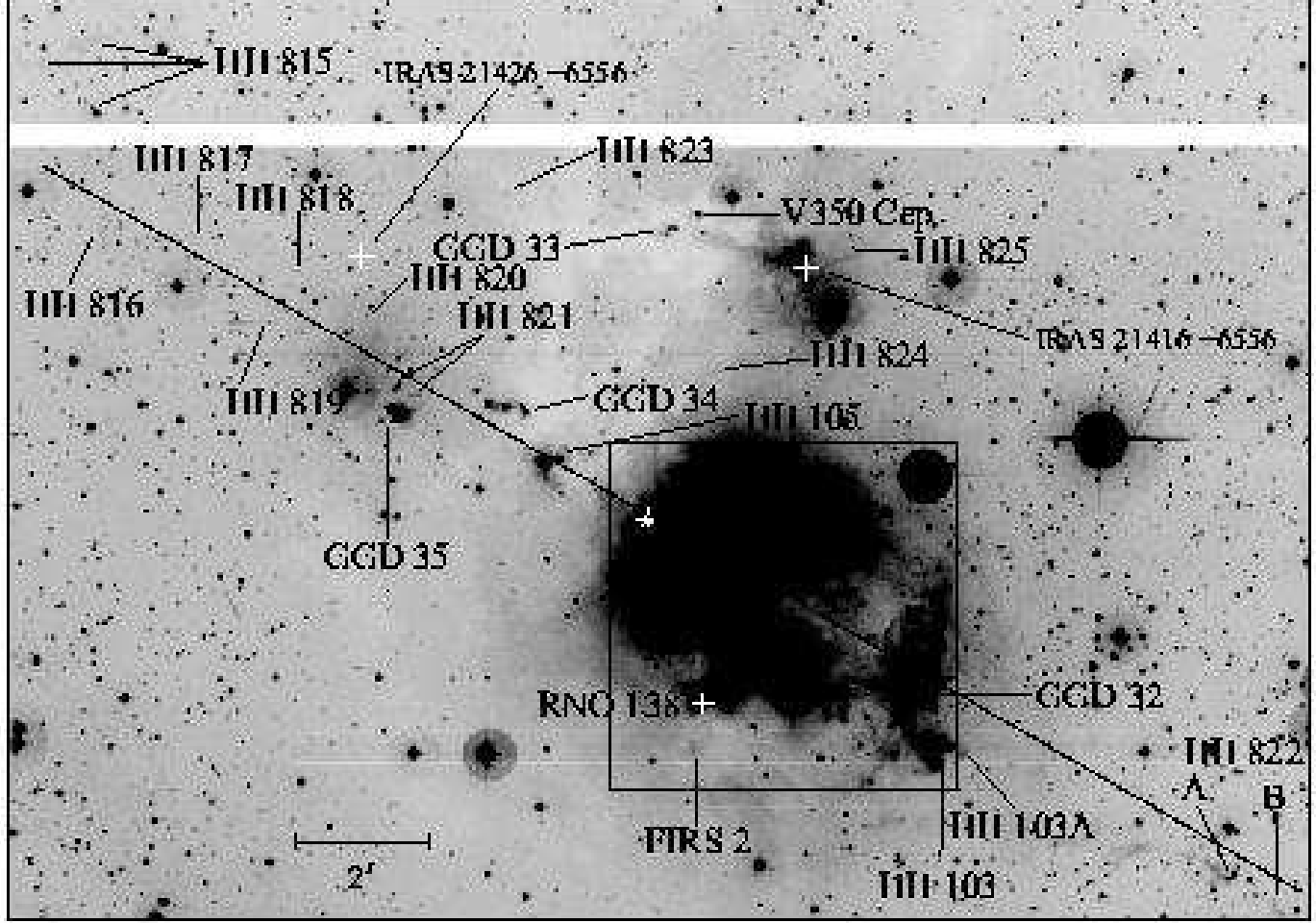}}
\caption{{\it LkH$\alpha$ 234 [SII]:\/} Mosaic of the entire outflow 
around \Lk$\!\!$. The position of \Lk is marked with a white star - 
the area surrounding \Lk indicated by the box can be seen more clearly in 
Fig.\ \ref{LkHa234centre}. The dotted line marks 
the primary outflow axis flow at 60\degr$\!$\,/\,240\degr$\!$. 
There is an optically invisible IRAS source in the cluster, IRAS\,21418+6552,
the position of which is marked with a white cross in Fig.\ \ref{LkHa234centre}. 
Other known infrared sources in the region are also indicated.}
\label{LkHa234flow}
\end{figure*}

\clearpage

\begin{figure}
\resizebox{\hsize}{!}{\includegraphics{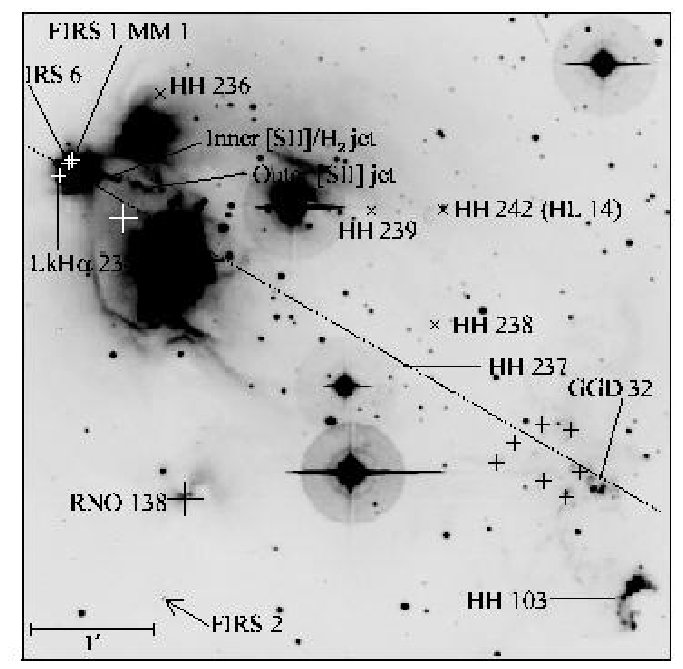}}
\caption{{\it LkH$\alpha$ 234 [SII] :\/} The cluster region  
(indicated in Fig.\ \ref{LkHa234flow}) including the  
``inner'' and ``outer'' optical jets \citep{Ray90,Cabrit97}. The 
contrast has been changed here with respect to Fig.\ \ref{LkHa234flow} so that 
more objects within the cluster core are visible. The white cross to the southwest
of \Lk marks the position of IRAS\,21814+6552. 
Black crosses and X's are used to mark the positions of HH objects found
by \cite{Eiroa92} and \cite{Miranda93} respectively.
The apparent primary outflow axis through the cluster at 
60\degr$\!$\,/\,240\degr\ is marked.}
\label{LkHa234centre}
\end{figure}

\clearpage

\begin{figure}
\resizebox{\hsize}{!}{\includegraphics{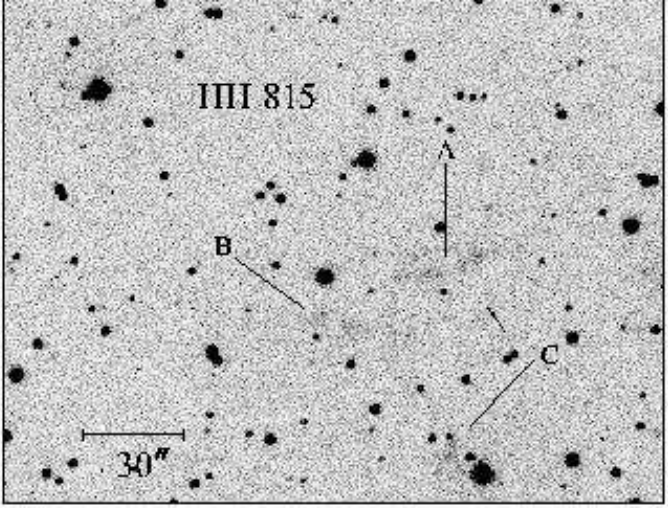}}
\caption{{\it LkH$\alpha$ 234 [SII]:\/} HH\,815, to the northeast of the 
cluster, is the most distant, known, outlying HH object in this region. Its 
relative position is seen in Fig.\ \ref{LkHa234flow}.}
\label{LkHa234NE2}
\end{figure}

\clearpage

\clearpage
 
\end{document}